\documentclass[twocolumn,showpacs,preprintnumbers]{revtex4}
\usepackage{amssymb}

\usepackage[dvips]{graphicx}

\begin{document}

\title{Unconventional critical behavior of fermions hybridized with bosons}
\author{A.S.~Alexandrov}
\affiliation{Department of Physics, Loughborough University,
Loughborough LE11 3TU, United Kingdom}

\begin{abstract}

A phase transition into the condensed state of fermions hybridized
with immobile bosons  is examined beyond the ordinary mean-field
approximation (MFA) in two and three dimensions. The hybridization
interaction does not provide the Cooper pairing of fermions and
the Bose-condensation in  two-dimensions. In the three-dimensional
(3D) boson-fermion model (BFM) an expansion in the strength of the
order parameter near the transition  yields no linear homogeneous
term in the Ginzburg-Landau-Gor'kov equation. This  indicates that
previous mean-field discussions of the model are flawed in any
dimensions. In particular,  the conventional (MFA) upper critical
field is zero in any-dimensional BFM
\end{abstract}
\pacs{PACS:  71.20.-z,74.20.Mn, 74.20.Rp, 74.25.Dw}

 \maketitle

\bigskip
Soon after Anderson \cite{pand} and Street and Mott \cite{mot}
introduced localized electron pairs to explain some unusual
properties of chalcogenide glasses, a two component model of
negative $U$ centers coupled with the Fermi sea of itinerant
fermions was  employed to study superconductivity in disordered
metal-semiconductor alloys \cite{sim,tin}. When the attractive
potential $U$ is large, the model is reduced to localized
hard-core bosons spontaneously decaying into itinerant electrons
and vice versa, different from a non-converting mixture of mobile
charged bosons and fermions \cite{ale3,and}. Later on the model
was applied more generally to describe pairing electron processes
with localization-delocalization \cite{ion},  and  a linear
resistivity in the normal state of cuprates \cite{eli}. The model
attracted more attention in connection with high-temperature
superconductors \cite{lee,ran,ran0,lar,ale,dam,dam2}. In
particular, Refs. \cite{dam,dam2} claimed  that 2D BFM with
immobile hard-core bosons is capable to reproduce some physical
properties and the phase diagram of cuprates.

The model is defined by the Hamiltonian,
\begin{eqnarray}
H &=&\sum_{{\bf k},\sigma =\uparrow ,\downarrow }\xi _{{\bf k}}c_{{\bf k}%
,\sigma }^{\dagger }c_{{\bf k},\sigma }+E_{0}\sum_{{\bf q}}b_{{\bf q}%
}^{\dagger }b_{{\bf q}}+ \\
&&{\rm g}N^{-1/2}\sum_{{\bf q,k}}\left( \phi _{{\bf k}}b_{{\bf q}}^{\dagger
}c_{-{\bf k}+{\bf q/2},\uparrow }c_{{\bf k}+{\bf q/2},\downarrow
}+H.c.\right) ,  \nonumber
\end{eqnarray}
where $\xi _{{\bf k}}=-2t(\cos k_{x}+\cos k_{y})-\mu $ is the 2D energy
spectrum of fermions, $E_{0}\equiv \Delta _{B}-2\mu $ is the bare boson
energy with respect to the chemical potential $\mu $, ${\rm g}$ is the
magnitude of the anisotropic hybridization interaction, $\phi _{{\bf k}%
}=\phi _{-{\bf k}}$ is the anisotropy factor, and $N$ is the number of
cells. Ref. \cite{dam} argued that 'superconductivity is induced in this
model from the anisotropic charge exchange (hybridization) interaction (${\rm g}\phi _{{\bf k%
}}$) between the conduction-band fermions and the immobile
hard-core bosons', and 'the on-site Coulomb repulsion  competes
with this pairing' reducing the critical temperature $T_{c}$ less
than by 25\%. Also it has been argued \cite{dam2}, that the
calculated upper critical field of the model fits well the
experimental results.

This, as well as some other  studies of BFM   applied a
 mean-field approximation (MFA) to the condensed phase of BFM, replacing  zero-momentum boson operators by c-numbers
 and neglecting  the boson self-energy.
MFA led to a conclusion
 that 'bosons exist only as virtual state' at sufficiently large  boson energy $E_{0}$, so that
 BFM exhibits features compatible with BCS characteristics with a relatively small fluctuation region
 $Gi$\cite{lar},
  and describes
 a crossover from the BCS-like to local pair  behavior \cite{lee,ran,dam}.
However,
 our study of BFM \cite{ale}  beyond MFA
revealed a crucial effect of the boson self-energy on the normal
state boson spectral function. The energy of zero-momentum bosons
is renormalized down to \emph{zero}  at the critical temperature
$T=T_c$, no matter how weak the boson-fermion coupling and how
large the bare boson energy are. As a result,  the Cooper pairing
of fermions via \emph{virtual} \emph{unoccupied } bosonic states
is impossible, but  it could occur only simultaneously with the
Bose-Einstein condensation of \emph{real} bosons in  BFM.

 Here I show that there is no BCS-like condensed
 state in the two-dimensional model.
  More surprisingly,  the mean-field
approximation appears meaningless even in three-dimensional BFM
because of the complete boson softening. The phase
 transition is never a BCS-like second-order phase transition. In
 particular, the conventional upper critical field is
 zero  in 3D BFM.

Replacing boson operators by $c$-numbers for ${\bf q}=0$ in Eq.(1)
one obtains  a linearised BCS-like equation for the fermion
order-parameter $\Delta _{{\bf k}}$,
\begin{equation}
\Delta _{{\bf k}}=\frac{{\rm \tilde{g}}^{2}\phi _{{\bf k}}}{E_{0}N}\sum_{%
{\bf k}^{\prime }}\phi _{{\bf k}^{\prime }}{\frac{\Delta _{{\bf k}^{\prime
}}\tanh (\xi _{{\bf k}^{\prime }}/2k_{B}T)}{{2\xi _{{\bf k}^{\prime }}}},%
}
\end{equation}
with the coupling constant  ${\rm \tilde{g}}^{2}={\rm
g}^{2}(1-2n^{B})$, renormalized by the hard-core effects. Using a
two-particle fermion vertex part in the Cooper channel one can
prove that this equation is perfectly correct even beyond the
conventional non-crossing approximation \cite{ale}. Nevertheless,
the problem with MFA stems from an incorrect definition of the
bare boson energy with respect to the chemical potential,
$E_{0}(T)$. This energy is determined by the atomic density of
bosons ($n^{B}$)  as (Eq.(9) in Ref. \cite{dam})
\begin{equation}
\tanh \frac{E_{0}}{2k_{B}T}=1-2n^{B}.
\end{equation}
While Eq.(2) is correct, Eq.(3) is incorrect because the boson
self-energy $\Sigma _{b}({\bf q},\Omega _{n})$ due to the same
hybridization  interaction is missing. At first sight \cite{dam}
the self-energy  is small in comparison to the kinetic energy of
fermions, if ${\rm g}$ is small. However $\Sigma _{b}(0,0)$
diverges logarithmically at zero temperature \cite{ale}, no matter
how week the interaction is. Therefore it should be kept in the
density sum-rule, Eq.(3). Introducing the boson Green's function
\begin{equation}
D({\bf q},\Omega _{n})=\frac{1-2n^{B}}{i\Omega _{n}-E_{0}-\Sigma _{b}({\bf q}%
,\Omega _{n})}
\end{equation}
one must replace incorrect Eq.(3) by

\bigskip
\begin{equation}
-{\frac{k_{B}T}{{N}}}\sum_{{\bf q},n}e^{i\Omega _{n}\tau }D({\bf q}%
,\Omega _{n})=n^{B},
\end{equation}
where $\tau =+0$, and $\Omega _{n}=2\pi k_{B}Tn$ ($n=0,\pm 1,\pm
2...$). The divergent (cooperon) contribution to $\Sigma _{b}({\bf
q},\Omega _{n})$ is given by \cite{ale}
\begin{eqnarray}
&&\Sigma _{b}({\bf q},\Omega _{n})=-\frac{{\rm \tilde{g}}^{2}}{2N}\sum_{{\bf %
k}}\phi _{{\bf k}}^{2}\times  \\
&&\frac{\tanh [\xi _{{\bf k-q/2}}/(2k_{B}T)]+\tanh [\xi _{{\bf k+q/2}%
}/(2k_{B}T)]}{\xi _{{\bf k-q/2}}+\xi _{{\bf k+q/2}}-i\Omega _{n}},
\nonumber
\end{eqnarray}
so that one obtains
\begin{equation}
\Sigma _{b}({\bf q},0)=\Sigma _{b}(0,0)+\frac{q^{2}}{2M^{\ast }}+{\cal O}%
(q^{4})
\end{equation}
for small ${\bf q}$ with any anisotropy factor compatible with the
point-group symmetry of the cuprates. Here $M^{\ast }$ is the boson mass,
calculated analytically in Ref.\cite{ale} with the isotropic exchange
interaction and parabolic fermion band dispersion (see also Ref.\cite{cris}%
), and $\hbar=1$. The BCS-like equation (2) has a nontrivial
solution for $\Delta _{{\bf k}} $ at $T=T_c$, if
\begin{equation}
E_{0}=-\Sigma _{b}(0,0).
\end{equation}
Substituting Eq.(7) and Eq.(8) into the sum-rule, Eq.(5) one obtains a
logarithmically divergent integral with respect to ${\bf q}$, and
\begin{equation}
T_{c}=\frac{const}{\int_{0}dq/q}=0.
\end{equation}

The devastating result, Eq.(9) is a direct consequence of the
well-known theorem, which states that BEC is impossible in 2D.
This is true for non-interacting bosons. Remarkably it appears
also true for bosons hybridized with fermions in 2D, Eq.(9).  Any
dynamic repulsion between bosons could provide an infrared cut-off
of the
 integral in Eq.(9) resulting in a finite critical temperature
 in 2D. But it has to be  treated beyond MFA
level \cite{pop,mic}, and in no way does it  make MFA
\cite{dam,dam2} a meaningful approximation.

One may erroneously believe that MFA results  can be still applied
in three-dimensions. However, increasing dimensionality does not
make MFA a meaningful approximation either. This approximation
leads to a naive conclusion that a BCS-like superconducting state
 occurs below the
 critical temperature   $T_{c}\simeq \mu \exp\left( -{%
E_{0}/z_c}\right) $ via fermion pairs being \emph{virtually}
excited into
 $unoccupied$ $virtual$ bosonic states \cite{lee,ran,lar}.  Here $z_c=\tilde{g}^{2}N(0)$ and
$N(0)$ is the density of states (DOS) in the fermionic band near
the Fermi level $\mu $. However,  the Cooper pairing of fermions
 is not possible via virtual unoccupied bosonic states  in 3D BFM either. Indeed,
Eq.(8) does not depend on the dimensionality, so that the
analytical continuation of Eq.(4) to real frequencies $\omega$
yields the partial boson DOS as $\rho(\omega)=(1-2n_B)
\delta(\omega)$ at $T=T_c$ for ${\bf q}=0$ in any-dimensional BFM
and for any coupling with fermions. Hence, the Cooper pairing may
occur only simultaneously with the Bose-Einstein condensation of
real bosons in  3D BFM \cite{ale}. The origin of the simultaneous
condensation of the fermionic and bosonic fields in 3D BFM lies in
the complete softening of the boson mode at $T=T_c$ caused by its
hybridization with fermions.

Taking into account the boson damping and dispersion shows that
the boson spectrum significantly changes for all momenta.
Continuing the self-energy, Eq.(6) to real frequencies yields  the
damping (i.e. the imaginary part of the self-energy) as \cite{ale}
\begin{equation}
\gamma({\bf q},\omega)={\pi z_c\over{4q\xi}} \ln
\left[{\cosh(q\xi+\omega/(4k_{B}T_{c}))\over{\cosh(-q\xi+\omega/(4k_{B}T_{c}))}}\right],
\end{equation}
where $\xi=v_F/(4k_{B}T_{c})$ is a coherence length. The damping
is significant when $q\xi<<1$. In this region $\gamma({\bf
q},\omega)=\omega\pi z_{c}/(8k_{B}T_{c})$ is comparable or even
larger than  the boson energy $\omega$. Hence bosons look like
overdamped diffusive modes, rather than quasiparticles in the
long-wave limit \cite{ale,cris}, contrary to the erroneous
conclusion of Ref.\cite{ran0}, that there is 'the onset of
coherent free-particle-like motion of the bosons' in this limit.
Only outside the long-wave   region, the damping becomes small.
Indeed, using Eq.(10) one obtains $\gamma({\bf q},\omega)=\omega
\pi z_{c}/(2qv_F)<< \omega$, so that bosons at
 $q >>1/\xi$ are well defined quasiparticles
 with a logarithmic dispersion, $\omega(q)=z_c \ln(q
\xi)$ \cite{ale}.  As a result the bosons disperse over the whole
energy interval from zero up to $E_0$.

The main mathematical problem with MFA stems from the density sum
rule, Eq.(5) which determines the chemical potential of the system
and consequently the bare boson energy $E_{0}(T)$ as a function of
temperature. In the framework of MFA one takes the bare boson
energy   in Eq.(2) as a temperature independent parameter,
$E_0=z_c\ln (\mu/T_c)$ \cite{lar}, or determines it from the
conservation of the total number of particles, Eq.(3) neglecting
the boson self-energy $\Sigma_b({\bf q}, \Omega)$
\cite{lee,ran,dam,dam2}). Then Eq.(2) looks like a conventional
Ginzburg-Landau-Gor'kov equation \cite{gor} linearized near the
transition with a negative coefficient $\alpha \propto T-T_c<0$ at
$T<T_c$,
\begin{equation}
\alpha \Delta({\bf r})=0,
\end{equation}
where
\begin{equation}
\alpha = 1+{\Sigma_b(0,0)\over{E_0}}\approx 1-
{g^2N(0)\over{E_0}}\ln {\mu\over{T}}.
\end{equation}
 Hence, one
concludes that the phase transition is almost a conventional
BCS-like transition, at least at $E_0\gg T_c$ \cite{lee,ran,lar}.
Also, using the Gor'kov expansion in powers of $\Delta$ in the
external field, one finds a finite upper critical field
$H_{c2}(T)$ \cite{dam2}.

However, these findings are mathematically and physically flawed.
Indeed, the term of the sum in Eq.(5) with $\Omega_n=0$ is given
by the integral
\begin{equation}
T\int {d{\bf q}\over{2\pi^3}}{1\over{E_0+\Sigma_b({\bf q},0)}}.
\end{equation}
 The integral converges, if and
only if $E_0\geqslant -\Sigma_b(0,0)$. Hence the coefficient
$\alpha(T)$ in Eq.(11) can not be negative at any temperature
below $T_c$ contrary to the MFA result \cite{lar,dam2}, which
 violates the density
sum-rule predicting a wrong negative $\alpha(T)$.

 Since $\alpha(T)\geqslant 0$,
  the  phase
 transition is never a BCS-like second-order phase transition
 even at large $E_0$ and small $g$. In fact, the
 transition  is driven by the Bose-Einstein condensation of \emph{
 real} bosons with ${\bf q}=0$, which occur  due to the complete
 softening of their spectrum at  $T_c$ in 3D BFM.
 Remarkably, the conventional upper critical field, determined as the field, where a non-trivial
 solution of the linearised Gor'kov equation \cite{gor} in the external field  occurs, is
 zero in 3D BFM, $H_{c2}(T)=0$, because $\alpha(T)\geqslant 0$ below $T_c$ (for more details see \cite{alecon}).
 It is not  a finite $H_{c2}(T)$
 found in Ref. \cite{dam2} using MFA. Of course, like in  2D BFM,
the  dynamic repulsion  between bosons could provide a finite
$H_{c2}(T)$ in 3D BFM, similar to the case of intrinsically mobile
3D bosons \cite{aleH}. However,  to provide a finite $H_{c2}(T)$
 the dynamic repulsion has to be treated beyond MFA, as discussed in Ref. \cite{aleH}.  Even
 at  temperatures well below $T_c$ the condensed state is fundamentally
 different from the BCS-like MFA ground state, because of the \emph{pairing} of
 bosons \cite{alecon}. It is similar  to a pairing of supracondensate helium atoms in $^4He$,
 proposed as an explanation for a low density of  the single-particle condensate\cite{pas}.
The boson pairing  appears in 3D BFM due to the hybridization of
bosons  with fermionic condensate. It is not expected in the
framework of MFA, where the effective interaction between bosons
was found  repulsive by integrating out the fermionic degrees of
freedom with an incorrect (constant) $E_0$ \cite{lar}. The
pair-boson condensate  should significantly modify the
thermodynamic properties of the condensed BFM compared with the
MFA predictions.

This qualitative failure of MFA  might be rather unexpected, if
one believes that bosons in Eq.(1) play the same role as phonons
in the BCS superconductor. This is not the case for two reasons.
The first one is  the density sum-rule, Eq.(5), for bosons which
is not applied to phonons. The second being that the boson
self-energy is given by the divergent (at $T=0$) Cooperon diagram,
while the self-energy of phonons is finite at small coupling.

I  conclude that  MFA results for the boson-fermion model  do make
any sense neither in two nor in three dimensions, because the
divergent self-energy has been neglected in calculating $T_{c}$
and $H_{c2}(T)$. Any  repulsion between bosons, induced by
 hard-core effects and/or by hybridization with fermions  could
not make MFA results \cite{lee,ran,ran0,lar,dam,dam2} meaningful.
It is well known that the "infrared-save" 2D theory by Popov and
others \cite{pop,mic} is actually \emph{not} a mean-field theory.
In 3D BFM the phase transition is due to the Bose condensation of
real bosons rather than virtual ones. Hence, the mean-field theory
cannot be applied for a description of   3D BFM either.  There is
no BCS-like state in any-dimensional BFM, and no BCS to local pair
crossover. The common wisdom that at weak coupling the
boson-fermion model is adequately described by the BCS theory, is
negated by our results.

I highly appreciate enlightening   discussions with A.F. Andreev,
 V.V. Kabanov,   A.P. Levanyuk, R. Micnas, and S.
Robaszkiewicz,  and the support by the Leverhulme Trust (UK) via
Grant F/00261/H.

\end{document}